\documentclass{article}

\usepackage{microtype}
\usepackage{graphicx}
\usepackage{subcaption}
\usepackage{booktabs} 

\usepackage{hyperref}

\usepackage[preprint]{icml2026}

\usepackage{amsmath}
\usepackage{amssymb}
\usepackage{mathtools}
\usepackage{amsthm}

\usepackage[capitalize,noabbrev]{cleveref}

\theoremstyle{plain}

\theoremstyle{definition}

\theoremstyle{remark}

\usepackage[textsize=tiny]{todonotes}

\icmltitlerunning{PianoKontext: Expressive Performance Rendering from Deadpan Context}

\begin{document}

\twocolumn[
  \icmltitle{PianoKontext: Expressive Performance Rendering from Deadpan Context}



  \icmlsetsymbol{equal}{*}

  \begin{icmlauthorlist}
    \icmlauthor{Dmitrii Gavrilev}{yyy}
  \end{icmlauthorlist}

  \icmlaffiliation{yyy}{Applied AI Institute}

  \icmlcorrespondingauthor{Dmitrii Gavrilev}{dmitrygavrilyev@gmail.com}

  \icmlkeywords{Machine Learning, ICML}

  \vskip 0.3in
]



\printAffiliationsAndNotice{}  

\begin{abstract}
Expressive performance rendering (EPR) aims to generate realistic performances constrained on sequences of notes. However, flow matching audio editing models manipulate only synchronized music samples of the same duration, limiting their understanding of expressive timing. We introduce PianoKontext, a flow matching rendering model for classical piano music that generates variable-length performances in the latent space of a pretrained Music2Latent model. We synthesize MIDI scores into deadpan audio and employ Dynamic Time Warping (DTW) in the latent space to construct paired data for training. The aligned embeddings are concatenated in DiT blocks, allowing for a simple and effective learning of the dependencies between the score and performances. Audio samples are available at our demo page: \url{https://realfolkcode.github.io/pianokontext_demo/}.
\end{abstract}

\section{Introduction}
Controllable music generation aims to bridge the gap between human artistic vision and generative models. A particular example is editing sketches of music or expressive rendering of note sequences. Although the research field of machine learning for music has been rapidly improving, there are still open problems that limit the usage of these models. First, deep learning methods for music editing predominantly focus on tasks that modify pairs of musical samples with the same duration (e.g., timbre transfer), paying less attention to expressive timing. Second, representing polyphonic music involves complex instrument-specific tokenization schemes \cite{miditok2021} or image-like piano roll arrays that are computationally expensive to denoise \cite{min2023polyffusion}. Moreover, rendering in the symbolic domain requires rigorous note-level alignment between scores and performances, which complicates modeling ambiguous phrasing, such as grace notes, trills, and other ornamentation \cite{borovik2026pianocore}. On the other hand, unlike symbolic music models, audio models might not be completely faithful to the score, hallucinating or omitting notes.

In this work, we propose PianoKontext, a latent flow matching model that renders a variable-length piano performance segments given a deadpan latent context. Our framework is inspired by FLUX Kontext \cite{labs2025flux}, an image editing model that jointly models the dependencies between the given context and target images with self-attention in Diffusion Transformer (DiT) blocks \cite{peebles2023scalable}. Similarly, we synthesize score MIDI into audio with a simple soundfont to serve as a deadpan context. We introduce a data preprocessing step based on Dynamic Time Warping (DTW) \cite{sakoe1970similarity} that enables an effective sampling of paired score-performance data. PianoKontext has an improved audio fidelity and lower hallucination rate compared to an unsupervised inversion baseline, while producing temporally consistent samples with different predefined durations. Our design is agnostic of instrument and can be readily extended to other music genres.

\section{Related Work}

\begin{figure*}[t]
  \vskip 0.2in
  \begin{center}
    \centerline{\includegraphics[width=2\columnwidth]{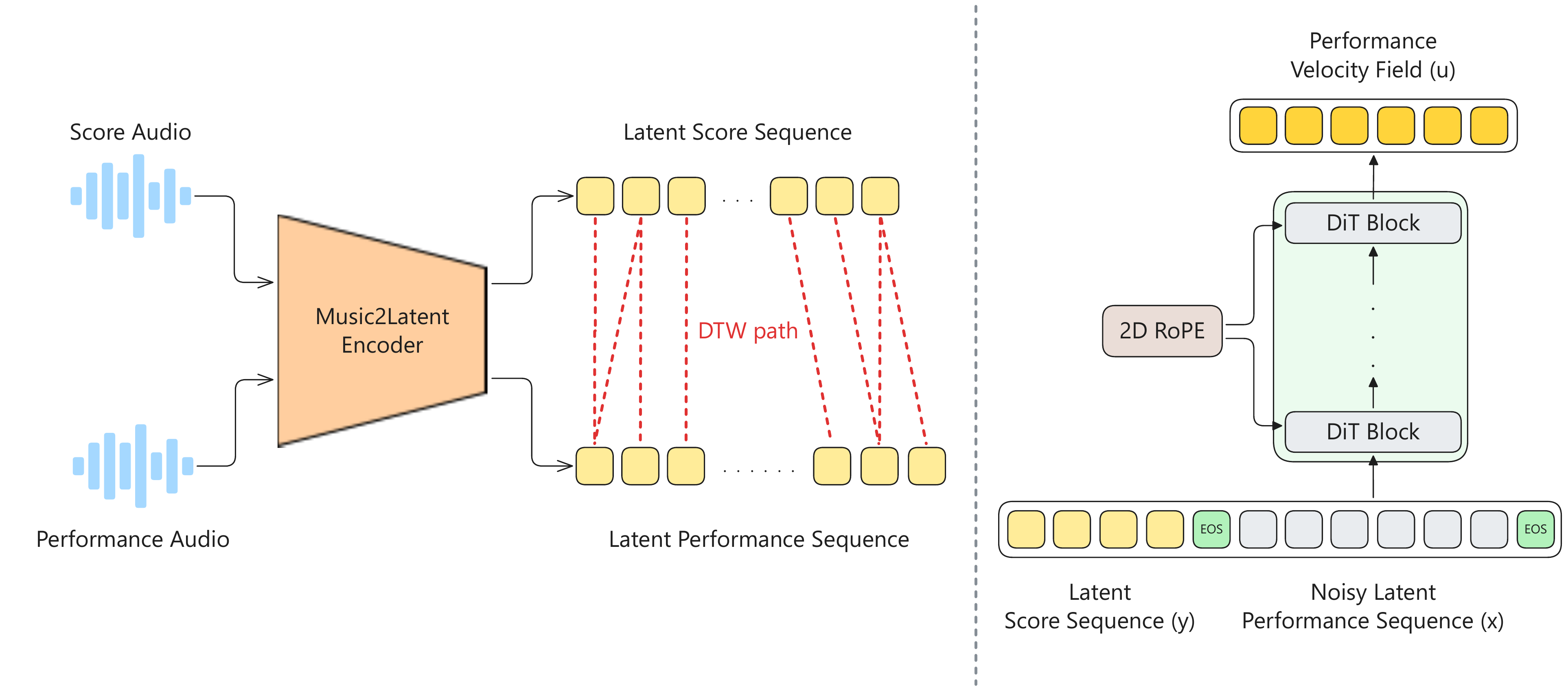}}
    \caption{Overview of PianoKontext. (Left) Preprocessing: Score and performance audiofiles are encoded with the pretrained Music2Latent model. The produced embeddings are then aligned with the DTW algorithm. (Right) Architecture: PianoKontext uses a concatenated score, noise, and EOS latents as its inputs, which are then passed to DiT blocks with 2D RoPE embeddings.
    }
    \label{pianokontext}
  \end{center}
\end{figure*}

\subsection{Latent Music Models}
Music generation has seen a surge of works focusing on modeling in the latent space of audio compression models, leveraging compact and lightweight representations for training. Autoregressive models, such as MusicGen \cite{copet2023simple}, decode discrete tokens from the quantized vocabulary of a neural audio codec \cite{defossezhigh}. Another paradigm relies on modeling the embeddings from continuous autoencoders \cite{pasinimusic2latent, evans2025stable} with diffusion or flow matching \cite{ho2020denoising, lipmanflow}.

Iterative denoising of latents enables flexible design of controllable generation through techniques such as classifier-free guidance. An example of an audio translation problem is timbre transfer, where diffusion models are applied to retrieve a semantic code of the source audio by inversion in the latent space of Music2Latent \cite{mancusi2025latent, lee2026diffusion}.

\subsection{Expressive Performance Rendering}
The research field of EPR has seen a growing interest in approaching the problem using deep learning methods. It has two major branches: 1) modeling expressive parameters in MIDI format, and 2) synthesizing performances directly in the audio domain. The former subfield aims to model attributes of each individual note, e.g., timing, velocity, and articulation. However, symbolic music models are agnostic to the acoustic properties of an instrument and space, which limits their application. 

On the other hand, end-to-end training of EPR models in the latent or the audio domain remains underresearched. One of the examples of such models is RenderBox \cite{zhang2025renderbox}, which is a finetuned Stable Audio Open model \cite{evans2025stable} that takes a MIDI score and a text prompt as controls. Instead of MIDI, GuitarFlow \cite{loth2025guitarflow} takes synthesized deadpan audio as its input, bypassing the need to learn a symbolic music encoder. Nevertheless, it can be trained only on perfectly aligned music examples, which hinders its ability to model expressive timing.

\section{Method}

\subsection{Background}
\label{sec:background}
In what follows, we briefly describe the key concepts behind flow matching \cite{lipmanflow}, a generative modeling paradigm that continuously interpolates between noise and data distributions $p_0$ and $p_1$, respectively. The intermediate states $x_t$ follow the marginal distribution $p_t$, $t \in [0, 1]$, and are constructed to linearly interpolate between the endpoints:
\begin{equation}
\label{eq:interpolation}
    x_t = \left( 1 - t \right) x_0 + t x_1.
\end{equation}
Then, there exists a velocity field $u_t(x_t)$ that induces an ordinary differential equation (ODE) with noise as its initial condition such that the solution $x_1 = x_0 + \int_0^1 u_t(x_t) dt$ adheres to the data distribution. In practice, regressing $u_t$ is intractable; nevertheless, it can be learned by regressing the conditional velocity field instead:
\begin{equation}
    u_t(x | x_1) = \frac{x_1 - x}{1 - t}.
\end{equation}

Optimizing the following objective (conditional flow matching loss) yields the approximation of the marginal velocity:
\begin{equation}
    \mathcal{L}_{\textrm{CFM}}(\theta) = \mathbb{E}_{t, x_1, x_t} \| u_t^{\theta}(x_t) - u_t(x_t | x_1) \|^2,
\end{equation}
where $u^{\theta}_t$ is the velocity parametrized by a neural network.

\subsection{PianoKontext}

The overall data pipeline and architecture are illustrated in Figure \ref{pianokontext}. First, we encode raw score and performance audios with the pretrained Music2Latent encoder. The produced sequence embeddings have a much lower sampling rate $\approx$ 11 Hz, and each element has a dimensionality of 64. In the next step, we calculate the alignment between the latent score and performance sequences with DTW. The purpose of this step is to enable sampling segments of latents that correspond to the same musical content. Since DTW is precomputed only once, it does not add computational overhead to the training process. 

Given a context deadpan audio $y$, the goal is to generate an expressive performance $x$ that preserves the musical content of $y$. Both $x$ and $y$ are latent sequences from the pretrained Music2Latent model. Thus, we frame the problem as audio-to-audio translation in the latent space, aiming to learn a conditional distribution $p(x | y)$. We employ guided flow matching, which extends the training method described in Section \ref{sec:background} with conditioning on the context variables \cite{lipman2024flow}.

To construct input for training, we first draw a score-performance pair and its corresponding DTW path. We then sample a random subpath from the precomputed DTW path such that the segment lengths do not exceed the maximum predefined sequence length $S$. Note that in practice, both segments in minibatches should have random lengths to ensure independence. Additionally, we impose a lower bound to the DTW subpath length to mitigate extremely short samples. The drawn segments are temporally aligned, share musical content, and differ in duration (e.g., a deadpan score may have slower tempo than a performance or vice versa). Next, we inject Gaussian noise into a target performance $x$ according to the schedule in Eq. \eqref{eq:interpolation}. We append $y$ and $x$ with learnable end-of-sequence (EOS) embeddings to reinforce temporal consistency due to variable-length sequences. Finally, both segments are concatenated to form input $(y,x)$.

We draw inspiration from FLUX Kontext \cite{labs2025flux} and implement DiT enriched with 2D Rotary Position Embeddings (RoPE). Unlike the original RoPE for language modeling \cite{su2024roformer}, 2D RoPE was introduced to encode relative positional information across multiple axes \cite{heo2024rotary}. We introduce an additional axis that separates performance elements from context. Specifically, the element position is encoded as $(i, s)$, where $i \in \{ 0, 1 \}$ is a binary indicator for context/performance elements, and $s \in \{ 0, 1, \dots, S \}$ denotes temporal position. Then, self-attention in DiT blocks jointly models $(y, x)$, capturing dependencies between scores and performances. The model outputs the performance velocity field, and the context elements are discarded.

\section{Experiments}

\begin{table}[t]
  \centering
  \begin{tabular}{|l|c|c|c|}
    \hline
    Split & \# Scores & \# Performances &\# Hours \\
    \hline
    Train & 118 & 399 & 34.21 \\
    \hline
    Validation & 21 & 46 & 3.34 \\
    \hline
    Test & 39 & 74 & 5.91 \\
    \hline
  \end{tabular}
  \vspace{0.2cm}
  \caption{The statistics of the aligned piano dataset used in our experiments.}
  \label{tab:datasets}
  \vspace{-5mm}
\end{table}

\subsection{Data}
We construct a paired dataset of classical Western piano music using two sources. Expressive performances are derived from MAESTRO \cite{hawthorne2018enabling}, a corpus of high-quality 44.1-48 kHz performed by virtuoso pianists. We synthesize deadpan context MIDI samples from the ASAP dataset \cite{Peter-2023} into audio using a YDP Grand Piano soundfont from the FreePats project. ASAP provides the scores for a subset of MAESTRO performances, making the construction of a paired deadpan-expressive dataset feasible. Table \ref{tab:datasets} presents the overall dataset statistics across the splits.

\subsection{Baseline}

We compare PianoKontext with an unsupervised baseline model that is based on trajectory inversion, which is a popular method for steering flow and diffusion models while preserving the content. For timbre transfer, \cite{mancusi2025latent} propose Dual Bridge, which is a composition of two diffusion models trained independently on different instrument-specific datasets. First, the source latent is inverted to the noise domain with the source model. Then, the obtained noise is decoded with the target model.

For our baseline, we take a similar approach but train a single conditional model on the combined ASAP and MAESTRO datasets. We assign a label to each sample ("deadpan" and "expressive" for ASAP and MAESTRO, respectively), which is then used to guide the model with classifier-free guidance \cite{ho2021classifier}. The inversion is performed with the source label, whereas denoising uses the target expressive label. We coin this method "CFG Bridge".

\subsection{Training details}

\begin{table*}[t]
\centering
\resizebox{0.75\textwidth}{!}{%
\begin{tabular}{@{}lccccc@{}}
\toprule
& FAD ($\downarrow$) & KAD ($\downarrow$) & Pitch DTW ($\uparrow$) & Alignment Precision ($\uparrow$) & Alignment Recall ($\uparrow$) \\
\midrule
CFG Bridge & 4.69 & 1.68 & 0.856 & 0.466 & 0.373 \\
PianoKontext & \textbf{2.96} & \textbf{0.91} & \textbf{0.888} & 0.630 & 0.666 \\
Human & - & - & 0.883 & \textbf{0.829} & \textbf{0.794} \\
\bottomrule
\end{tabular}%
}
\vspace{0.2cm}
\caption{Evaluation metrics.}
\vspace{-5mm}
\label{tab:metrics}
\end{table*}

\begin{figure*}[t]
     \centering
     \begin{subfigure}{0.33\textwidth}
         \centering
         \includegraphics[width=\textwidth]{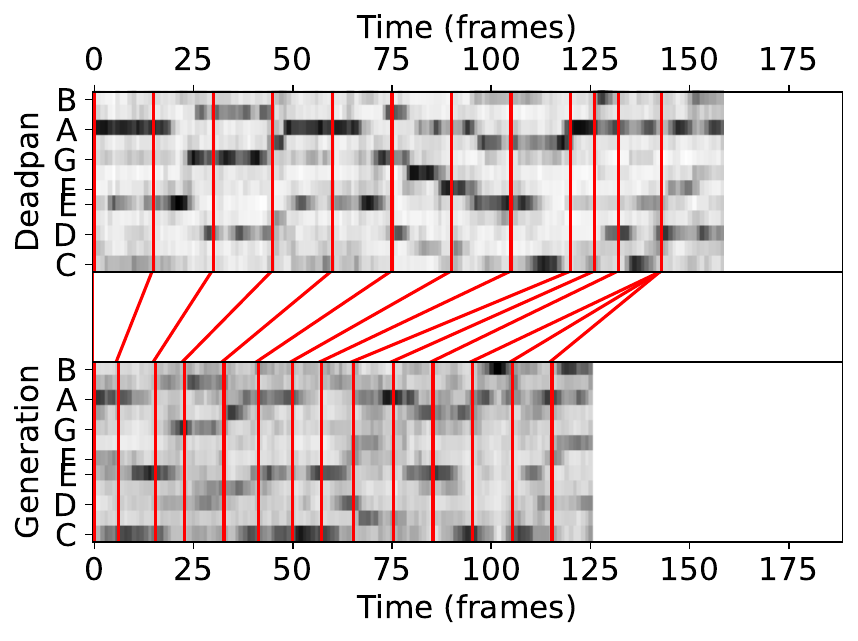}
         \caption{Duration factor = 0.8}
         \label{fig:duration_08}
     \end{subfigure}
     \hfill
     \begin{subfigure}{0.33\textwidth}
         \centering
         \includegraphics[width=\textwidth]{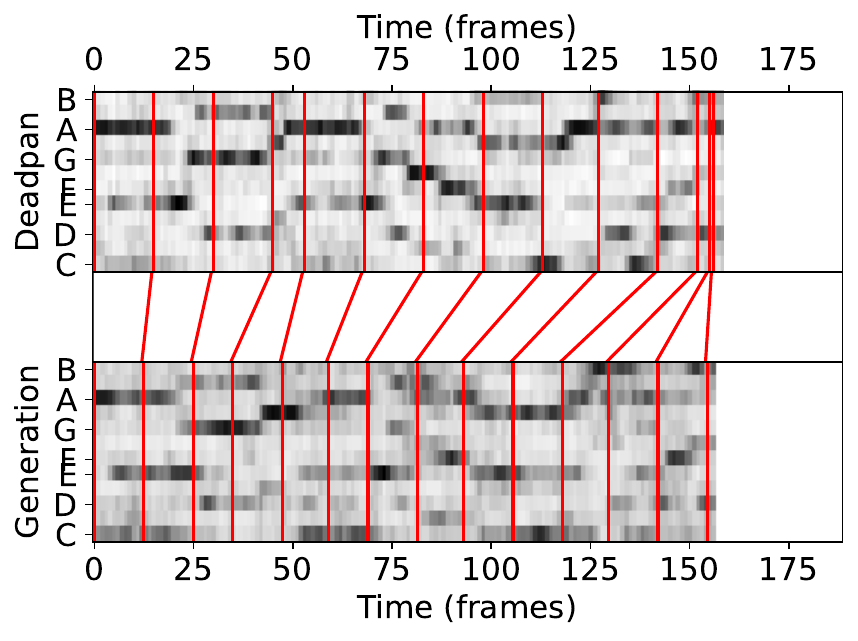}
         \caption{Duration factor = 1}
         \label{fig:duration_10}
     \end{subfigure}
     \hfill
     \begin{subfigure}{0.33\textwidth}
         \centering
         \includegraphics[width=\textwidth]{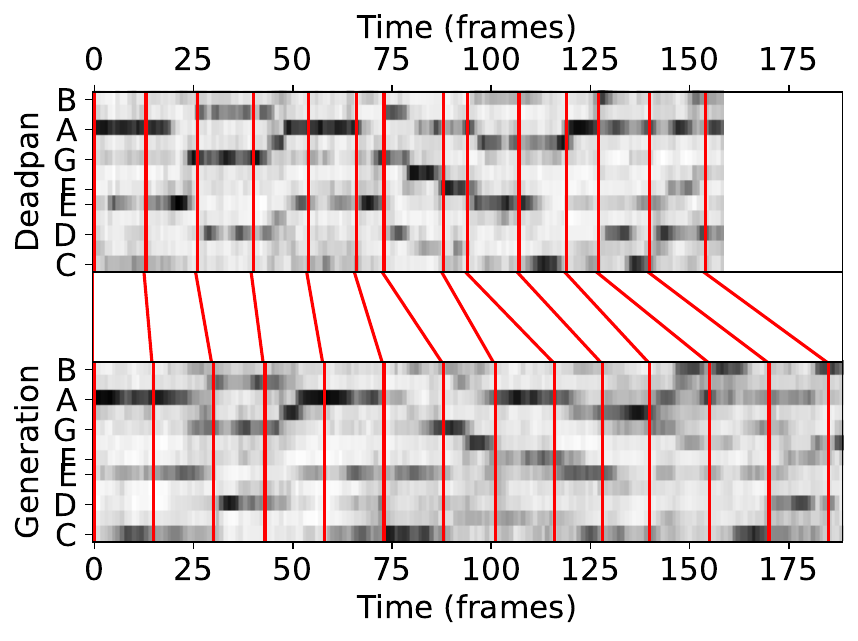}
         \caption{Duration factor = 1.2}
         \label{fig:duration_12}
     \end{subfigure}
        \caption{An example of PianoKontext inference with different predefined durations. (Top) Synthesized deadpan score. (Bottom) Generated performances. The red lines indicate DTW paths.}
        \label{fig:duration}
    \end{figure*}

Both CFG Bridge and PianoKontext share the same DiT architecture: 8 DiT blocks with a hidden size of 512 and an MLP expansion ratio of 1. Each DiT block in CFG Bridge has 16 attention heads, whereas we halve this number for PianoKontext to increase the per-head dimensionality due to an additional RoPE axis.

We calculate the shared dataset statistics for deadpan and expressive latents and standardize the dataset before training. For training, we set the maximum sequence length $S = 128$, which corresponds to 11 s of audio. CFG Bridge and PianoKontext are trained on a single NVIDIA RTX 4090 GPU for 240k and 120k iterations, respectively. We use AdamW optimizer \cite{LoshchilovH19} with cosine scheduler for learning rate (base lr = 5e-4 and weight decay = 0.01), batch size = 128, and Exponential Moving Average. CFG Bridge is trained with $10\%$ of class dropout.

\subsection{Evaluation}

We generate five performances for each score from the test set, using only the first few seconds. For PianoKontext, we sample deadpan context and human performances of different latent lengths that do not exceed $S$. Since CFG Bridge operates on sequences of fixed size, we employ the first 11s of context. Performances are generated using 64 steps of the Heun ODE solver. CFG Bridge uses different guidance scales for inversion (1.0) and denoising (2.0).

To evaluate the audio fidelity of the samples, we employ Frechet Audio Distance (FAD) \cite{gui2024adapting} and Kernel Audio Distance (KAD) \cite{kad}, audio generation metrics that measure the distributional discrepancies between the embedding sets. The embeddings are extracted using MERT-95M \cite{li2024mert}. In addition, we measure the DTW cosine similarity between the CQT chromagrams of the deadpan and generated performances \cite{zhang2025renderbox}. We also adopt Alignment Recall and Precision from \cite{borovik2026pianocore}. To this end, we transcribe generated audio performances with Transkun, a state-of-the-art piano transcription model \cite{yan2024scoring}. The transcribed MIDI are then aligned with the groundtruth scores using Parangonar \cite{peter-offline2023}. Then the constructed note alignment is used to quantify the ratios of missing and hallucinated notes.

\section{Results and Discussion}

Table \ref{tab:metrics} presents the results of the evaluation. PianoKontext outperforms an unsupervised bridge method in audio fidelity (FAD, KAD) and content preservation (Pitch DTW, Alignment Precision, Alignment Recall). Human-level Pitch DTW indicates the preservation of structure and harmony and can be seen as a "soft" metric. The precision and recall, which are "hard" due to their discrete nature, suggest that PianoKontext is significantly less prone to deviations from the score compared to CFG Bridge. Since the transcription model is not perfect, we also provide the metrics for human performances that serve as upper bounds.

We showcase an example of performances synthesized with PianoKontext in Figure \ref{fig:duration}.We provide the 7‑second opening of Debussy’s “Pour le piano” as context. The noise sequence is initialized with different duration factors. That is, we choose the factors 0.8, 1, and 1.2 relative to the context length to generate performances with varying tempos. The chromagram features indicate the distribution of pitch classes. We also plot the DTW paths to visualize the alignment between the score and generated samples. PianoKontext successfully generates performances in different tempos, highlighting the controllability of our framework. The opening features 16th‑note arpeggios marked \emph{non legato}, indicating clarity of each note. Although PianoKontext follows the structure, harmony, and melody, it lacks the desired articulation. Nevertheless, we encourage the reader to listen to the audio samples provided on our demo page.

In this paper, we proposed a proof-of-concept for learning expressive polyphonic music conditioned on deadpan context sequences. Future directions involve exploring how adoption of other instruments affects the model's understanding of expressivity, as well as a more rigorous evaluation of musicality. Extending the sequence length and the incorporation of outpainting techniques could facilitate generation of full-length performances.

\section*{Acknowledgements}
The author would like to thank Ilya Borovik, Jackson Loth, Vladimir Viro, and Dmitry Yarotsky for advice and fruitful discussion.



\bibliography{example_paper}

@inproceedings{
  hawthorne2018enabling,
  title={Enabling Factorized Piano Music Modeling and Generation with the {MAESTRO} Dataset},
  author={Curtis Hawthorne and Andriy Stasyuk and Adam Roberts and Ian Simon and Cheng-Zhi Anna Huang and Sander Dieleman and Erich Elsen and Jesse Engel and Douglas Eck},
  booktitle={International Conference on Learning Representations},
  year={2019},
  url={https://openreview.net/forum?id=r1lYRjC9F7},
}

@article{Peter-2023,
 title = {Automatic Note-Level Score-to-Performance Alignments in the ASAP Dataset},
 author = {Peter, Silvan David and Cancino-Chacón, Carlos Eduardo and Foscarin, Francesco and McLeod, Andrew Philip and Henkel, Florian and Karystinaios, Emmanouil and Widmer, Gerhard},
 doi = {10.5334/tismir.149},
 journal = {Transactions of the International Society for Music Information Retrieval {(TISMIR)}},
 year = {2023}
}

@article{zhang2025renderbox,
  title={Renderbox: Expressive performance rendering with text control},
  author={Zhang, Huan and Maezawa, Akira and Dixon, Simon},
  journal={arXiv preprint arXiv:2502.07711},
  year={2025}
}

@inproceedings{evans2025stable,
  title={Stable audio open},
  author={Evans, Zach and Parker, Julian D and Carr, CJ and Zukowski, Zack and Taylor, Josiah and Pons, Jordi},
  booktitle={ICASSP 2025-2025 IEEE International Conference on Acoustics, Speech and Signal Processing (ICASSP)},
  pages={1--5},
  year={2025},
  organization={IEEE}
}

@article{loth2025guitarflow,
  title={GuitarFlow: Realistic Electric Guitar Synthesis From Tablatures via Flow Matching and Style Transfer},
  author={Loth, Jackson and Sarmento, Pedro and Sandler, Mark and Barthet, Mathieu},
  journal={arXiv preprint arXiv:2510.21872},
  year={2025}
}

@article{copet2023simple,
  title={Simple and controllable music generation},
  author={Copet, Jade and Kreuk, Felix and Gat, Itai and Remez, Tal and Kant, David and Synnaeve, Gabriel and Adi, Yossi and D{\'e}fossez, Alexandre},
  journal={Advances in neural information processing systems},
  volume={36},
  pages={47704--47720},
  year={2023}
}

@article{defossezhigh,
  title={High Fidelity Neural Audio Compression},
  author={D{\'e}fossez, Alexandre and Copet, Jade and Synnaeve, Gabriel and Adi, Yossi},
  journal={Transactions on Machine Learning Research},
  year={2024}
}

@inproceedings{pasinimusic2latent,
  title={Music2Latent: Consistency Autoencoders for Latent Audio Compression},
  author={Pasini, Marco and Lattner, Stefan and Fazekas, George},
  booktitle={Proceedings of the International Society for Music Information Retrieval Conference, {ISMIR}},
  year = {2024}
}

@article{ho2020denoising,
  title={Denoising diffusion probabilistic models},
  author={Ho, Jonathan and Jain, Ajay and Abbeel, Pieter},
  journal={Advances in neural information processing systems},
  volume={33},
  pages={6840--6851},
  year={2020}
}

@inproceedings{lipmanflow,
  title={Flow Matching for Generative Modeling},
  author={Lipman, Yaron and Chen, Ricky TQ and Ben-Hamu, Heli and Nickel, Maximilian and Le, Matthew},
  booktitle={The Eleventh International Conference on Learning Representations},
  year={2023}
}

@inproceedings{mancusi2025latent,
  title={Latent diffusion bridges for unsupervised musical audio timbre transfer},
  author={Mancusi, Michele and Halychanskyi, Yurii and Cheuk, Kin Wai and Moliner, Eloi and Lai, Chieh-Hsin and Uhlich, Stefan and Koo, Junghyun and Mart{\'\i}nez-Ram{\'\i}rez, Marco A and Liao, Wei-Hsiang and Fabbro, Giorgio and others},
  booktitle={ICASSP 2025-2025 IEEE International Conference on Acoustics, Speech and Signal Processing (ICASSP)},
  pages={1--5},
  year={2025},
  organization={IEEE}
}

@article{lee2026diffusion,
  title={Diffusion Timbre Transfer Via Mutual Information Guided Inpainting},
  author={Lee, Ching Ho and Nistal, Javier and Lattner, Stefan and Pasini, Marco and Fazekas, George},
  journal={arXiv preprint arXiv:2601.01294},
  year={2026}
}

@article{labs2025flux,
  title={FLUX. 1 Kontext: Flow Matching for In-Context Image Generation and Editing in Latent Space},
  author={Labs, Black Forest and Batifol, Stephen and Blattmann, Andreas and Boesel, Frederic and Consul, Saksham and Diagne, Cyril and Dockhorn, Tim and English, Jack and English, Zion and Esser, Patrick and others},
  journal={arXiv preprint arXiv:2506.15742},
  year={2025}
}

@article{kad,
    author={Chung, Yoonjin and Eu, Pilsun and Lee, Junwon and Choi, Keunwoo and Nam, Juhan and Chon, Ben Sangbae},
    title={KAD: No More FAD! An Effective and Efficient Evaluation Metric for Audio Generation}, 
    journal = {arXiv:2502.15602},
    url = {https://arxiv.org/abs/2502.15602},
    year = {2025}
}

@inproceedings{gui2024adapting,
  title={Adapting frechet audio distance for generative music evaluation},
  author={Gui, Azalea and Gamper, Hannes and Braun, Sebastian and Emmanouilidou, Dimitra},
  booktitle={ICASSP 2024-2024 IEEE International Conference on Acoustics, Speech and Signal Processing (ICASSP)},
  pages={1331--1335},
  year={2024},
  organization={IEEE}
}

@inproceedings{sakoe1970similarity,
  title={A similarity evaluation of speech patterns by dynamic programming},
  author={Sakoe, Hiroaki and Chiba, Seibi},
  booktitle={Nat. Meeting of Institute of Electronic Communications Engineers of Japan},
  volume={136},
  year={1970}
}

@article{borovik2026pianocore,
  title={PianoCoRe: Combined and Refined Piano MIDI Dataset},
  author={Borovik, Ilya},
  journal={Transactions of the International Society for Music Information Retrieval},
  volume={9},
  number={1},
  year={2026}
}

@inproceedings{yan2024scoring,
  author    = {Yujia Yan and Zhiyao Duan},
  title     = {Scoring Time Intervals Using Non-Hierarchical Transformer for Automatic Piano Transcription},
  booktitle = {Proc. International Society for Music Information Retrieval Conference (ISMIR)},
  year      = {2024},
}

@inproceedings{peter-offline2023,
  title={Online Symbolic Music Alignment with Offline Reinforcement Learning},
  author={Peter, Silvan David},
  booktitle={International Society for Music Information Retrieval Conference {(ISMIR)}},
  year={2023}
}

@inproceedings{min2023polyffusion,
  title={Polyffusion: A Diffusion Model for Polyphonic Score Generation With Internal and External Controls},
  author={Min, Lejun and Jiang, Junyan and Xia, Gus and Zhao, Jingwei},
  booktitle={Ismir 2023 Hybrid Conference},
  year={2023}
}

@inproceedings{miditok2021,
    title={{MidiTok}: A Python package for {MIDI} file tokenization},
    author={Fradet, Nathan and Briot, Jean-Pierre and Chhel, Fabien and El Fallah Seghrouchni, Amal and Gutowski, Nicolas},
    booktitle={Extended Abstracts for the Late-Breaking Demo Session of the 22nd International Society for Music Information Retrieval Conference},
    year={2021},
    url={https://archives.ismir.net/ismir2021/latebreaking/000005.pdf},
}

@inproceedings{li2024mert,
  title={Mert: Acoustic music understanding model with large-scale self-supervised training},
  author={Li, Yizhi and Yuan, Ruibin and Zhang, Ge and Ma, Yinghao and Chen, Xingran and Yin, Hanzhi and Xiao, Chenghao and Lin, Chenghua and Ragni, Anton and Benetos, Emmanouil and others},
  booktitle={International Conference on Learning Representations},
  volume={2024},
  pages={12181--12204},
  year={2024}
}

@article{lipman2024flow,
  title={Flow matching guide and code},
  author={Lipman, Yaron and Havasi, Marton and Holderrieth, Peter and Shaul, Neta and Le, Matt and Karrer, Brian and Chen, Ricky TQ and Lopez-Paz, David and Ben-Hamu, Heli and Gat, Itai},
  journal={arXiv preprint arXiv:2412.06264},
  year={2024}
}

@inproceedings{LoshchilovH19,
  author       = {Ilya Loshchilov and
                  Frank Hutter},
  title        = {Decoupled Weight Decay Regularization},
  booktitle    = {7th International Conference on Learning Representations, {ICLR} 2019,
                  New Orleans, LA, USA, May 6-9, 2019},
  publisher    = {OpenReview.net},
  year         = {2019},
  url          = {https://openreview.net/forum?id=Bkg6RiCqY7},
  bibsource    = {dblp computer science bibliography, https://dblp.org}
}

@inproceedings{peebles2023scalable,
  title={Scalable diffusion models with transformers},
  author={Peebles, William and Xie, Saining},
  booktitle={Proceedings of the IEEE/CVF international conference on computer vision},
  pages={4195--4205},
  year={2023}
}

@article{su2024roformer,
  title={Roformer: Enhanced transformer with rotary position embedding},
  author={Su, Jianlin and Ahmed, Murtadha and Lu, Yu and Pan, Shengfeng and Bo, Wen and Liu, Yunfeng},
  journal={Neurocomputing},
  volume={568},
  pages={127063},
  year={2024},
  publisher={Elsevier}
}

@inproceedings{heo2024rotary,
  title={Rotary position embedding for vision transformer},
  author={Heo, Byeongho and Park, Song and Han, Dongyoon and Yun, Sangdoo},
  booktitle={European Conference on Computer Vision},
  pages={289--305},
  year={2024},
  organization={Springer}
}

@inproceedings{ho2021classifier,
  title={Classifier-Free Diffusion Guidance},
  author={Ho, Jonathan and Salimans, Tim},
  booktitle={NeurIPS 2021 Workshop on Deep Generative Models and Downstream Applications},
  year={2021}
}
\bibliographystyle{icml2026}



\end{document}